\documentclass[letterpaper, 10 pt, conference]{ieeeconf}  

\usepackage{amstext}

\IEEEoverridecommandlockouts                              
\title{\LARGE \bf
Information-Theoretic Security in Wireless Networks\thanks{This
work was prepared under the support of the U.S. National Science
Foundation under Grants ANI-03-38807 and CNS-06-25637. } }

\author {Yingbin Liang, H. Vincent Poor\thanks{Y. Liang and H. V. Poor are with
the Department of Electrical Engineering, Princeton University,
Princeton, NJ 08544 USA {\tt \{yingbinl,poor\}@princeton.edu}.}
 and Shlomo Shamai (Shitz)\thanks{S. Shamai (Shitz) is with the
Department of Electrical Engineering, Technion-Israel Institute of
Technology, Technion City, Haifa 32000, Israel {\tt
sshlomo@ee.technion.ac.il}. }}

\begin{document}

\maketitle
\thispagestyle{empty}
\pagestyle{empty}


\section*{SUMMARY}

\noindent Security in wireless networks has traditionally been
considered to be an application layer issue. However, with the
emergence of ad hoc and other less centralized networking
environments, there has been an increase in interest in the
potential of the wireless physical layer to provide communications
security. Information theory provides a natural framework for the
study of this issue, and consequently there has been a resurgence
of interest in information-theoretic security in recent years,
particularly for wireless channel models. Of course, the use of
information theoretic concepts to characterize communications
security dates to Shannon's earliest work, and the important work
on the wire-tap channel by Wyner and by Csisz\'{a}r and
K$\ddot{\text{o}}$rner in the 1970's addressed security issues for
broadcast communications. But, recent work has taken these early
ideas and expanded on them considerably, by examining multiple
access channels, cognitive communications, compound channels, fading,
multiple-antenna  (MIMO) transmission, code design for secure transmission,
feedback, authentication, secure network coding, and many other
issues. 

This paper will review recent contributions of the authors
and their co-workers in this general area. In particular, a new channel
model for wireless multiple access communication will be
introduced. For this channel, in which security issues arise
naturally, recent results characterizing reliable communication
rates under secrecy constraints (possibly perfect secrecy) will be
presented. Then, general design principles that are essential to
achieve secure communication will be postulated. Finally, additional recent
research on the topic of information-theoretic security will be
discussed briefly as time permits. These results include the
secrecy capacity region of fading broadcast channels, coding
design for wire-tap channels, and the impact of feedback on
secrecy rate. 

The talk will touch on the work of a number of
colleagues and co-workers, including Hesham El Gamal (Ohio State
University), Jared Grubb (University of Texas, Austin), Gerhard
Kramer (Alcatel-Lucent Bell Labs), Lifeng Lai (Princeton), Ruoheng
Liu (Princeton), Pedrag Spasojevic (Rutgers WINLAB),
 Anelia Somekh-Baruch (Princeton), Xiaojun Tang
(Rutgers WINLAB), Sergio Verd\'{u} (Princeton) and Sriram
Vishwanath (University of Texas, Austin), the details of which can
be found in the publications cited in the references.



\end{document}